\documentclass[12pt]{iopart}

\usepackage{times}
\usepackage{iopams}
\usepackage{graphicx}

\begin{document}

% You should use BibTeX and apsrev.bst for references
%\bibliographystyle{h-elsevier2}
%\preprint{}

\title{No sliding in time}
% Optional argument for running titles on pages
%\title[]{}



\author{Kirill Shtengel\dag, Chetan Nayak\ddag, Waheb Bishara\dag 
and Claudio Chamon\S}
%\email[]{shtengel@microsoft.com}
%\affiliation
%\institute{\inst{1}California Institute of Technology,
%Pasadena, CA 91125\\
%\inst{2} Dept. of Physics, UCLA\\
%\inst{3} Dept. of Physics, Boston University}
%\pacs{nn.mm.xx}{First pacs description}
\address{\dag\ California Institute of Technology, Pasadena, CA 91125}
\address{\ddag\ Department of Physics, UCLA, Los Angeles, CA 90095}
\address{\S\ Department of Physics, Boston University, Boston, MA
  02215} 
\ead{kirill@caltech.edu}

%\thanks{}
%\date{\today}
%\begin{document}

%\maketitle

\begin{abstract}
  In this letter, we analyse the following apparent paradox: As has
  been recently proved by Hastings \cite{Hastings04a}, under a general
  set of conditions, if a \emph{local} Hamiltonian has a spectral gap
  above its (unique) ground state (GS), all connected equal-time
  correlation functions of local operators decay exponentially with
  distance. On the other hand, statistical mechanics provides us with
  examples of 3D models displaying so-called sliding phases
  \cite{OHern99} which are characterised by the algebraic decay of
  correlations within 2D layers and exponential decay in the third
  direction. Interpreting this third direction as time would imply a
  gap in the corresponding (2+1)D quantum Hamiltonian which would
  seemingly contradict Hastings' theorem. The resolution of this
  paradox lies in the non-locality of such
  a quantum Hamiltonian.
\end{abstract}

%\pacs{}
%\maketitle must follow title, authors, abstract and \pacs
%\maketitle

\section{Introduction}
When discussing the properties of a quantum many-body Hamiltonian, one
commonly intermixes the notions of \emph{gaplessness} and
\emph{criticality}.  These two properties, however, may or may not
automatically imply each other.  The former refers to the absence of
an energy gap separating the GS(s) from the excited states and implies
slower than exponential, typically power-law decay of correlations in
imaginary time. The latter refers to the equal-time correlations of
operators separated in space. Naturally, if the system is
Lorentz-invariant, the two properties are identical. Even in the
absence of Lorentz invariance, these two properties typically follow
from each other.  \emph{E.g.}, at a quantum critical point, there is a
dynamical critical exponent $z$ relating the scaling of correlations
in space and in time: $t\propto x^z$. Since all standard examples of
quantum critical points are characterised by finite non-zero values of
$z$, gaplessness once again implies criticality and vise versa.
However, a spin model introduced recently in the context of systems
with topological order \cite{Freedman05b}, was shown to be gapless
while possessing only short-ranged equal time correlations between
\emph{local} operators. Moreover, in that particular model one could
add another term to the Hamiltonian which would open a gap without
affecting equal-time correlations. This last feature, of course,
should not come as a surprise since for any model with short-ranged
ground-state correlations, one could just add to the Hamiltonian an
operator $(\mathbb{I}-\mathcal{P}_{0})$ where $\mathcal{P}_{0}$ is a
projector onto this ground state. Such a projector will necessarily
open a spectral gap while the quasi-local nature of ground state
correlations will almost certainly lead to the quasi-local nature of
the contributing terms in the Hamiltonian. It is the former feature,
namely the spectral gaplessness of a system with short-range
correlations that comes as a surprise \footnote{Notice that such a
  possibility is remarkable from the point of view of understanding
  \emph{glasses}: one of the puzzles about a generic glassy behaviour
  is their slow dynamics with no long spatial correlations.}. As
discussed in \cite{Freedman05b}, this is due to a ``bottle-neck''
quantum dynamics that does not allow for the efficient mixing of
different quantum states contributing to the ground state and thus
allows one to construct a ``twisted'' excited state (in the spirit of
Lieb-Schultz-Mattis theorem \cite{Lieb61,Hastings04a}) whose energy is
vanishing in the thermodynamic limit. A similar situation was
encountered in models \cite{Castelnovo05a} where the ``bottle-neck''
quantum dynamics, in addition to leading to energy gaps that are
exponentially small in domain sizes, prevents the systems from
reaching the true ground state when coupled to a quantum bath at zero
temperature, remaining in a state of ``quantum glassiness''
\cite{Chamon05}.

A natural question then arises: could we also find a system described
by a \emph{local} Hamiltonian with a spectral gap which nevertheless
displays power-law decay of equal-time correlations of \emph{local}
operators.  The condition of locality is important here: relaxing the
requirement of locality of a Hamiltonian leads to a trivial ``yes'' by
the means of adding a projector onto the critical ground state as
described in the previous paragraph. We also know examples of gapped
systems with quasi-long ranged correlations of non-local operators
such as the non-local order parameters in the quantum Hall effect
\cite{Girvin87,Zhang89,Read89a} and other topological phases
\cite{Freedman04a,Freedman05b}.  In short, the answer to the above
question is ``No'' due to the theorem proved by Hastings
\cite{Hastings04a}. This, however, contradicts our intuition drawn
from Statistical Mechanics where \emph{sliding} phases are known to
exist, as has been demonstrated in the context of stacks of 2D layers
of $XY$ spins with gradient coupling between the layers
\cite{OHern99}. The resulting \emph{sliding} phase is characterised by
an algebraic decay of in-plane correlations within the layer and an
exponential decay in the perpendicular direction.  Later, this
construction has been extended to the quantum systems of stripes
\cite{Emery00,Vishwanath01}, with the layers now representing 1D
stripes in space-time. Such phases were also found in dissipative
Josephson junction arrays \cite{Tewari05a,Tewari05b}.

In general, it is common in the study of quantum critical phenomena of
$d$-dimensional systems to relate the problem to $(d+1)$-dimensional
classical systems. This can be done generically, and perhaps one place
where a true difference between quantum systems in Euclidean time and
classical systems may arise is when a topological term exists and the
resulting classical action cannot be made real. Typically, in going
from a $d$-dimensional quantum system with local interactions and
dynamics to a $(d+1)$-dimensional classical system, one maps the local
quantum Hamiltonian density into a local classical Lagrangian density.
Hence, if the critical properties of this local classical system can
be understood, so would those of the quantum model.  Now, one could
turn the question around, and ask instead about a behaviour of a model
derived from a local classical Lagrangian. Here we would like to
address this question in a particularly interesting case - that of a
classical sliding phase.  A quantum version of this model would imply
critical correlations in space but not in time, if the direction
perpendicular to the layers is taken to be Euclidean time. Thus the
system is expected to be gapped and yet have critical correlations
in space, seemingly a violation of Hastings' theorem. The purpose of
this Letter is to examine this apparent contradiction.

\section{Sliding Phase} 
We now quickly review the sliding phase as described in
\cite{OHern99}. The \emph{classical} Hamiltonian consists of three
parts:
\begin{equation}
  \label{eq:sliding}
  {H} =
{H}_0 + {H}_\mathrm{g} + {H}_\mathrm{J}.
\end{equation}
Here ${H}_0$ is just a
sum over all layers of independent $XY$-Hamiltonians:
\begin{equation}
\label{HXY}
{H}_0 = {K \over 2} \sum_n \int d^2 r
\left[{\boldsymbol{\nabla}}_{\perp} \theta_n(\mathbf{r})
\right]^2,
\end{equation}
where $\mathbf{r} = (x,y,0)$ is a point in the $x$-$y$ plane and
${\boldsymbol{\nabla}}_{\perp}$ is the gradient operator acting
on these two coordinates. 

The second term in Eq.~(\ref{eq:sliding}) couples gradients of
$\theta_n$ in different layers:
\begin{equation}
\label{HXYint}
{H}_\mathrm{g} = {1 \over 2}
\sum_{n,m} \int d^2r~{U_m \over 2}
\left\{{\boldsymbol{\nabla}}_{\perp}
\left[ \theta_{n+m} (\mathbf{r}) -
\theta_{n}(\mathbf{r})\right]\right\}^2.
\end{equation}
Finally,  Josephson couplings between layers are added:
\begin{equation}
\label{Josephson}
{H}_\mathrm{J} = - V_\mathrm{J} \int d^2r \cos\left[
\theta_{n+p}(\mathbf{r}) - \theta_{n}(\mathbf{r})\right].
\end{equation}
Following \cite{OHern99}, we choose to consider only two-layer
couplings with $p$ being the distance between the coupled layers. For
the reasons that will become clear shortly, we will concentrate on the
next-nearest layer coupling, $p=2$. As follows from \cite{OHern99},
the physics of the sliding phase should not depend on this choice.

In the absence of the gradient couplings (second term in
Eq.~(\ref{eq:sliding})), this Hamiltonian is just that of a 3D
$XY$-model (strictly speaking, for $p=2$, it is two independent
interlaced $XY$-models). It might appear unorthodox that we keep the
cosine coupling between the layers while using its expanded form for
the in-plane couplings. The continuous in-plane limit is not an issue
here; we simply adopt it from \cite{OHern99}. This choice of couplings
is permitted keeping in mind the ultimate goal of constructing the
phase with critical intra-layer and exponential inter-layer
correlations.  In such a phase, one can ignore in-plane vortices but
must allow for between-the-planes ones. However, without the gradient
couplings, there is no such phase as the temperature at which layers
decouple turns out to be higher than the Kosterlitz-Thouless (KT)
temperature in a single layer.

On the other hand, in the absence of Josephson couplings ($V_J = 0$),
one obtains the \emph{ideal} sliding Hamiltonian ${H}_\mathrm{S}={H}_0
+ {H}_\mathrm{g}$, which is invariant with respect to
$\theta_n(\mathbf{r}) \rightarrow \theta_n(\mathbf{r}) + \psi_n$ for
any constant $\psi_n$. I.e., the energy is unchanged when angles in
different layers slide relative to one another by arbitrary amounts --
one can think of it as a ``reduced'' gauge symmetry with one gauge
choice per layer.  As a result, the angles in different layers are
uncorrelated.  The low-temperature phase is the \emph{ideal} sliding
phase characterised by 
$\cos[\theta_m ( \mathbf{r} ) - \theta_n ( 0)] \sim
\delta_{m,n} r^{-\eta}$ with
$\eta = {T/(2\pi \tilde{K})}$,
where $\tilde{K}$ is the renormalised in-plane coupling.  For the
simplest case of only nearest layer coupling $U_m = U \delta_{m,\pm
  1}$, the effective in-plane coupling is $\tilde{K}= K\sqrt{1+4U/K}$.

What constitutes the main result of \cite{OHern99}, is the fact that
in the presence of \emph{both} the gradient and Josephson couplings
between the layers, a careful choice of coupling constants may open a
``window of opportunity'' in temperature within which the layers
decouple before the spins in each layer completely disorder via a KT
transition. In other words, the inter-layer vortices proliferate
before the in-plane vortices unbind. This is the sliding phase which
we shall now discuss in the quantum context.

\section{Quantum Hamiltonian}
Our discussion of constructing the corresponding quantum Hamiltonian
will follow Kogut's review \cite{Kogut79}. If the classical
Hamiltonian $H$ consists of only intra-layer terms, $H^\prime
[\theta_n]$ and terms coupling neighbouring layers, $H^{\prime\prime}
[\theta_n,\theta_{n+1}]$, the corresponding partition function can be
written as a $Z = \mathrm{tr}\left(\hat{T}^N\right)$ where the
interlayer transfer matrix is defined as
\begin{equation}
%  \fl
  \langle
  \theta_{n+1}\left|\hat{T}\right|\theta_{n}\rangle =
  \exp\left\{-\beta\left(\frac{1}{2}H^\prime [\theta_n] +
      \frac{1}{2}H^\prime [\theta_{n+1}] + H^{\prime\prime}
      [\theta_n,\theta_{n+1}]\right)\right\}
  \label{eq:transfermatrix_def}
\end{equation}
and $N$ is the total number of layers. We now identify the $n$-th
layer with the imaginary time slice $\tau_n$ while $\tau_{n+1} =
\tau_n + \epsilon$. The quantum Hamiltonian $\hat{\mathcal{H}}$ is
then defined by $\langle \theta(\tau_n +
\epsilon)\left|\exp\left\{-\epsilon \hat{\mathcal{H}}
  \right\}\right|\theta(\tau_n)\rangle =\langle \theta(\tau_n +
\epsilon)\left|\hat{T}\right|\theta(\tau_n)\rangle$ in the limit of
$\epsilon \to 0$. Notice that the explicit form of a quantum
Hamiltonian may be difficult to obtain: the transfer matrix needs not
be in the form of an exponential of a simple operator expression. It
is, however, often easy to represent it as a product of such
exponentials, each corresponding to a term in the classical
Hamiltonian.

\subsection{Ideal sliding} 
Let us first look at the case of an ideal sliding Hamiltonian ($V_J =
0$). For simplicity, for now we also restrict ourselves to the
nearest-neighbour coupling: $U_m = U \delta_{m,\pm 1}$.  Denoting
$\theta = {\theta}_{n}={\theta}^{(i)}(\tau)$ and $\theta^{\prime} =
{\theta}_{n+1}={\theta}^{(i)}(\tau+\epsilon)$ we have:
\begin{equation}
  \label{eq:transfermatrix_ideal_sliding}
  \fl
  \langle \theta^{\prime}| \hat{T}_\mathrm{S}|\theta \rangle = 
  \exp \left[-\frac{\beta}{4} \int d^2 r \left( 
      K \left[{\boldsymbol{\nabla}} \theta(\mathbf{r})
      \right]^2
      +K \left[{\boldsymbol{\nabla}} \theta^{\prime}(\mathbf{r})
      \right]^2
      + 2{U} \left\{{\boldsymbol{\nabla}} 
        \left[ \theta^{\prime} (\mathbf{r})
          - \theta (\mathbf{r})\right]\right\}^2 \right) \right].
\end{equation}
Introducing the canonically conjugate quantised fields
$\hat{\theta}(\mathbf{r})$ and $\hat{\pi}(\mathbf{r})$ such that
$\left[ \hat{\pi}(\mathbf{r}),\hat{\theta}(\mathbf{r'})\right] = -i
\delta(\mathbf{r'}-\mathbf{r})$, we obtain:
\begin{eqnarray}
  \label{eq:transfer_oper_ideal_sliding}
   \hat{T}_\mathrm{S} & \propto & \exp \left\{- \frac{\beta K}{4}\int d^2 r 
     \left[{\boldsymbol{\nabla}} \hat{\theta}(\mathbf{r})
      \right]^2\right\} 
    \nonumber\\
    & \times &
    \exp \left\{\frac{1}{4\pi \beta U}\int d^2 r d^2 r'
      \hat{\pi}(\mathbf{r}) \hat{\pi}(\mathbf{r'})
      \ln\left|\mathbf{r}-\mathbf{r'}\right|
    \right\}
    \nonumber\\
    & \times &
    \exp \left\{- \frac{\beta K}{4}\int d^2 r 
     \left[{\boldsymbol{\nabla}} \hat{\theta}(\mathbf{r})
      \right]^2\right\}.
\end{eqnarray}
This can be explicitly verified by substituting the above expression
for $\hat{T}$ into Eq.~(\ref{eq:transfermatrix_ideal_sliding}). Notice
that the operators in the first and the third exponent of
Eq.~(\ref{eq:transfer_oper_ideal_sliding}) are diagonal in the
$\theta$ representation while for the second exponential we have:
\begin{eqnarray}
  \label{eq:sliding_marix_element}
  \fl
  \langle \theta^{\prime}|
  \exp \left\{g\int d^2 r  d^2 r'
    \hat{\pi}(\mathbf{r}) \hat{\pi}(\mathbf{r'})
    \ln\left|\mathbf{r}-\mathbf{r'}\right|
  \right\}
  |\theta \rangle 
  \nonumber\\
  \lo
  =  \int \mathcal{D}p \mathcal{D}p'   
  \langle \theta^{\prime}|p^{\prime} \rangle \langle p^{\prime}|
  \exp \left\{g\int d^2 r  d^2 r'
    \hat{\pi}(\mathbf{r}) \hat{\pi}(\mathbf{r'})
    \ln\left|\mathbf{r}-\mathbf{r'}\right|
  \right\}
  |p\rangle \langle p |\theta \rangle 
  \nonumber\\
  \lo
  =  \int \mathcal{D}p
  \exp \left\{\int d^2 r \left[ \mathrm{i} p(\mathbf{r})
      [{\theta}^{\prime}(\mathbf{r}) - \theta(\mathbf{r})]
      + g\int  d^2 r'
      p(\mathbf{r}) p(\mathbf{r'})
      \ln\left|\mathbf{r}-\mathbf{r'}\right| 
    \right]
  \right\}
  \nonumber\\
  \lo
  \propto  \int \mathcal{D}p_k
  \exp \left\{\int \frac{d^2 k}{(2\pi)^2} \left[ \mathrm{i} p_\mathbf{k}
      \left({\theta}^{\prime}_{-\mathbf{k}} - \theta_{-\mathbf{k}}\right)
      -\frac{2\pi g}{k^2} p_\mathbf{k} p_{-\mathbf{k}} 
    \right]
  \right\}
  \nonumber\\
  \lo
  \propto  
  \exp \left\{-\frac{1}{8\pi g}\int \frac{d^2 k}{(2\pi)^2} \; {k^2} 
    \left({\theta}^{\prime}_{\mathbf{k}} - \theta_{\mathbf{k}}\right)
    \left({\theta}^{\prime}_{-\mathbf{k}} - \theta_{-\mathbf{k}}\right)
  \right\}
  \nonumber\\
  \lo
  =   \exp \left(-\frac{1}{8\pi g}\int d^2 r
    \left\{{\boldsymbol{\nabla}}
      \left[ \theta^{\prime} (\mathbf{r})
        - \theta (\mathbf{r})\right]\right\}^2
  \right) .
\end{eqnarray}

Instead of writing the corresponding quantum Hamiltonian formally as
$\hat{\mathcal{H}}_\mathrm{S}= - (1/{\epsilon})\ln
\hat{T}_\mathrm{S}$, we can simplify it by going to the continuous
time limit $\epsilon \to 0$. We introduce the new coupling constants,
$g_{\theta}\equiv \frac{\beta K}{2 \epsilon}$ and $g_{\pi}\equiv
\frac{1}{4 \pi \beta U \epsilon}$ and require that they do not scale
with $\epsilon$, which implies the following scaling for the original
couplings: $K \sim \epsilon$, $U \sim \epsilon^{-1}$.  Then, up to the
corrections of order $\epsilon$,
\begin{equation}
  \label{eq:quant_sliding_ham}
  \hat{\mathcal{H}}_\mathrm{S}=   g_{\theta} \int d^2 r 
   \left[{\boldsymbol{\nabla}} \hat{\theta}(\mathbf{r})
      \right]^2 
    - g_{\pi}  \int d^2 r d^2 r'
     \hat{\pi}(\mathbf{r}) \hat{\pi}(\mathbf{r'})
      \ln\left|\mathbf{r}-\mathbf{r'}\right|.
\end{equation}

Notice that the implied scaling of $K$ and $U$ does not lead to any
problems with tuning the parameters to the values needed to reach the
desired sliding phase.  This is because the \emph{effective} coupling
governing the behaviour of the classical statistical mechanical model
is given by
\begin{equation}
  \label{eq:effective_coupling}
  \beta \tilde{K}= \beta K\sqrt{1+\frac{4U}{K}}= 
  2 \epsilon g_{\theta} \sqrt{1+ \frac{1}{2
      \pi g_{\theta} g_{\pi} \epsilon^{2}}} 
  \to \sqrt{\frac{2 g_{\theta}}{\pi  g_{\pi}}} 
  \quad \mathrm{as\ } {\epsilon \to 0}. 
\end{equation}
Thus, this model has a well-defined continuous time limit.  

However, the quantum version of the ideal sliding Hamiltonian contains
logarithmically long ranged interactions between momenta at different
points. As will be discussed below, Hastings' theorem does not apply
to such a Hamiltonian, hence no contradiction appears here.  We remark
on an interesting physical picture arising from the quantum
Hamiltonian (\ref{eq:quant_sliding_ham}) if we think of the
eigenvalues of $\hat{\pi}$ as ``charge'' or ``vorticity''. Notice that
due to the compactness of the conjugate variable $\theta$, the
eigenvalues of $\hat{\pi}$ are quantised in integer units (we used
$\hbar =1$), this is completely analogous to the quantisation of $L_z$
in quantum mechanics. Therefore the second term in the quantum
Hamiltonian (\ref{eq:quant_sliding_ham}) describes a classical 2D
Coulomb gas (or a gas of vortices) with the usual logarithmic
interaction.  We know that the collective mode (plasmon) in such gas
is gapped (unlike in the case of $1/r$ interactions). The first term
has no simple classical meaning in this language; it is responsible
for making the correlations quasi-long ranged. Indeed, as follows from
Eq.~(\ref{eq:effective_coupling}), unless $g_{\theta}>
g_{\theta}^{\ast} \equiv 2 g_{\pi}/ \pi$, the model is in its
``high-temperature'' phase with both a spectral gap and exponentially
decaying correlations.

One could argue, however, that an ideal sliding phase is
``pathological'' in the sense of having different time slices
completely uncorrelated.  In what follows we argue that considering
all terms in the classical Hamiltonian~(\ref{eq:sliding}) does not
alleviate the above problem of long-range interactions while bringing
new ones.

\subsection{Non-ideal sliding} 
One apparent problem arises immediately: according to \cite{OHern99},
in order to have a sliding phase we must have additional gradient
couplings between at least next-nearest layers. Since no time
derivatives of momenta are allowed to appear in a Hamiltonian, such
coupling seems to have no quantum analog.  This problem is, however,
easily circumvented by doubling the number of components of the field
$\theta$:
\begin{equation}
  \boldsymbol{\theta}(\mathbf{r},\tau_m)= 
\left({\theta}^{(1)}(\mathbf{r},\tau_m),
{\theta}^{(2)}(\mathbf{r},\tau_m)\right) 
\equiv  \left(\theta_{2m}(\mathbf{r}),
\theta_{2m+1}(\mathbf{r})\right). 
  \label{eq:twocomp}
\end{equation}
A single time slice is now represented by two layers. Due to the
nearest-layer interactions in the original classical Hamiltonian we
have now generated interactions between the two components of our
quantum field $\boldsymbol{\theta}$, but this situation is not
unusual: a non-linear $\sigma$-model provides a standard example of
such behaviour. It must now become clear why we have chosen $p=2$ in
Eq.~(\ref{Josephson}): the Josephson terms now couple only identical
components in the nearest time slices.

Keeping track of both components, however, brings unnecessary
complications. The problem with constructing a local quantum
Hamiltonian is apparent even if we forget about the two-component
nature of the field and concentrate only on a single component. In
what follows, ${\vartheta}$ will be used to represent either of the
two components, ${\theta}^{(1)}$ or ${\theta}^{(2)}$, and for
simplicity we will only consider the terms in the classical
Hamiltonian that do not mix them :
\begin{eqnarray}
%\fl
  \tilde H = \frac{1}{2}\sum_n \int d^2 r \left( 
    {K} \left[{\boldsymbol{\nabla}}_{\perp} \theta_n(\mathbf{r})
    \right]^2  +  {U_2} \left\{{\boldsymbol{\nabla}}_{\perp} \left[
        \theta_{n+2} (\mathbf{r}) -
        \theta_{n}(\mathbf{r})\right]\right\}^2 
  \right.
  \nonumber\\
  \qquad \qquad \qquad \qquad \qquad \qquad \qquad 
  \left. 
    - 2V_\mathrm{J}
    \cos\left[ \theta_{n+2}(\mathbf{r}) -
      \theta_{n}(\mathbf{r})\right]\right).
  \label{eq:reduced_Hamiltonian}
\end{eqnarray}
%(in \cite{OHern99} $U_2<0$, but this is not essential here).  
Denoting $\vartheta = {\theta}_{n}={\theta}^{(i)}(\tau)$ and
$\vartheta^{\prime} = {\theta}_{n+2}={\theta}^{(i)}(\tau+\epsilon)$ we
have, similarly to Eq.~(\ref{eq:transfermatrix_ideal_sliding}):

\begin{eqnarray}
  \label{eq:transfermatrix_sliding}
  \fl
  \langle \vartheta^{\prime}| \hat{\tilde T}|\vartheta \rangle = 
  \exp \left[-\frac{\beta}{4} \int d^2 r \left( 
      K \left[{\boldsymbol{\nabla}} \vartheta(\mathbf{r})
      \right]^2
      +K \left[{\boldsymbol{\nabla}} \vartheta^{\prime}(\mathbf{r})
      \right]^2
      + 2{U_2} \left\{{\boldsymbol{\nabla}} 
        \left[ \vartheta^{\prime} (\mathbf{r})
          - \vartheta (\mathbf{r})\right]\right\}^2
    \right. \right.
  \nonumber\\
  \qquad \qquad \qquad \qquad \qquad \qquad  
  \left.  \left. 
      - 4V_\mathrm{J}
      \cos\left[ \vartheta^{\prime}(\mathbf{r}) 
        - \vartheta (\mathbf{r})\right]
    \right) \right].
\end{eqnarray}
With the help of $\exp(a \cos \phi) = \sum_m I_{|m|}(a)
\exp(\mathrm{i} m a)$, where $I_n(x)$ is a modified Bessel function,
we obtain, similarly to Eq.~(\ref{eq:transfer_oper_ideal_sliding}):
\begin{eqnarray}
  \label{eq:transfer_oper_sliding}
  \fl
  \hat{\tilde T} \propto \exp \left\{- \frac{\beta K}{4}\int d^2 r 
    \left[{\boldsymbol{\nabla}} \hat{\theta}(\mathbf{r})
    \right]^2\right\} 
  \nonumber\\
  \lo
  \times
  \sum_{\{m(\mathbf{r})\}}
  \exp \Bigg\{\int d^2 r d^2 r'
  \Big(\frac{1}{4\pi \beta U_2}
  \left[\hat{\pi}(\mathbf{r})-m(\mathbf{r})\right]
  \left[\hat{\pi}(\mathbf{r'})-m(\mathbf{r'})\right]
  \ln\left|\mathbf{r}-\mathbf{r'}\right|
  \nonumber\\
  \qquad \qquad \qquad 
  +
  \ln I_{|m(\mathbf{r})|}(\beta V_{\mathrm{J}})
  \Big)
  \Bigg\}
  \nonumber\\
  \lo
  \times
  \exp \left\{- \frac{\beta K}{4}\int d^2 r 
    \left[{\boldsymbol{\nabla}} \hat{\theta}(\mathbf{r})
    \right]^2\right\}.
\end{eqnarray}

While obtaining the actual quantum Hamiltonian from
Eq.~(\ref{eq:transfer_oper_sliding}) is much more complicated than
from Eq.~(\ref{eq:transfer_oper_ideal_sliding}), the two cases share
the same important feature: a long-ranged logarithmic interaction
between momenta at different locations.

\section{Discussion}
Let us now discuss the applicability of Hastings' theorem to the
quantum Hamiltonian we have attempted to construct. In short, the
theorem states that for a quantum system with a \emph{local}
Hamiltonian and a unique ground state separated from excited states by
a gap, all equal-time connected correlation functions of local
operators $\langle 0 | \hat{A} \hat{B} | 0 \rangle - \langle 0 |
\hat{A} | 0 \rangle\langle 0 |\hat{B} | 0 \rangle$ decay exponentially
with distance.

Therefore this theorem is not applicable to the quantum Hamiltonians
(\ref{eq:transfer_oper_ideal_sliding},\ref{eq:transfer_oper_sliding})
due to logarithmic interactions appearing there.  What seems
counterintuitive is that these non-local interactions originate from
a perfectly local ``sliding'' term in the classical action.

The resolution of the original paradox appears to be one aspect of,
perhaps, a broader question: when can one go from a {\it local} classical
Lagrangian to a {\it local} quantum Hamiltonian. As our example shows,
this needs not always be the case. 

We finally remark that locality can sometimes be restored at a cost of
introducing auxiliary degrees of freedom.  \emph{E.g.}, a ``Coulomb
gas'' analogy for the ideal sliding Hamiltonian
(\ref{eq:quant_sliding_ham}) readily hints at such a possibility: an
introduction of an auxiliary gauge field -- an electromagnetic vector
potential -- will make the theory local.  However, this will not
contradict Hastings' theorem, as such field will come with its own
gapless mode -- a photon!

%\acknowledgments
%\begin{acknowledgments}
\ack The authors are grateful to S.~Kivelson, S.~Chakravarty,
A.~Kitaev, I.~Dimov, X.-G.~Wen, E.~Fradkin, G.~Refael and I.~Klich for
useful discussions and comments. K.~S.\ and C.~N.\ have been supported by
the ARO under Grant No.~W911NF-04-1-0236.
C.~N.\ has also been supported by the NSF under
Grant No.~DMR-0411800.
%\end{acknowledgments}

\section*{References}
%\References
%\bibliography{../bibs/corr,../bibs/reference}

\begin{thebibliography}{10}

\bibitem{Hastings04a}
M.B. Hastings,
\newblock Phys. Rev. B 69 (2004) 104431, cond-mat/0305505.

\bibitem{OHern99}
C.S. O'Hern, T.C. Lubensky and J. Toner,
\newblock Phys. Rev. Lett. 83 (1999) 2745, cond-mat/9904415.

\bibitem{Freedman05b}
M. Freedman, C. Nayak and K. Shtengel,
\newblock Phys. Rev. Lett. 94 (2005) 147205, cond-mat/0408257.

\bibitem{Lieb61}
E. Lieb, T. Schultz and D. Mattis,
\newblock Ann. Phys. 16 (1961) 407.

\bibitem{Castelnovo05a}
C. Castelnovo et~al.,
\newblock (2004), cond-mat/0410562.

\bibitem{Chamon05}
C. Chamon,
\newblock Phys. Rev. Lett 94 (2005) 040402, cond-mat/0404182.

\bibitem{Girvin87}
S.M. Girvin and A.H. MacDonald,
\newblock Phys. Rev. Lett. 58 (1987) 1252.

\bibitem{Zhang89}
S.C. Zhang, T.H. Hansson and S. Kivelson,
\newblock Phys. Rev. Lett. 62 (1989) 82.

\bibitem{Read89a}
N. Read,
\newblock Phys. Rev. Lett. 62 (1989) 86.

\bibitem{Freedman04a}
M. Freedman et~al.,
\newblock Ann. Phys. 310 (2004) 428, cond-mat/0307511.

\bibitem{Emery00}
V.J. Emery et~al.,
\newblock Phys. Rev. Lett. 85 (2000) 2160, cond-mat/0001077.

\bibitem{Vishwanath01}
A. Vishwanath and D. Carpentier,
\newblock Phys. Rev. Lett 86 (2001) 676, cond-mat/0003036.

\bibitem{Tewari05a}
S. Tewari, J. Toner and S. Chakravarty,
\newblock (2004), cond-mat/0407308.

\bibitem{Tewari05b}
S. Tewari, J. Toner and S. Chakravarty,
\newblock (2005), cond-mat/0501219.

\bibitem{Kogut79}
J.B. Kogut,
\newblock Rev. Mod. Phys. 51 (1979) 659.

\end{thebibliography}

\end{document}